# Floquet topological insulator laser

Sergey K. Ivanov,[1,2,3] Yiqi Zhang,[1,*] Yaroslav V. Kartashov,[3] and Dmitry V. Skryabin[4]

[1]*Department of Applied Physics, School of Science, Xi'an Jiaotong University, Xi'an 710049, China*

[2]*Moscow Institute of Physics and Technology, Dolgoprudny, Moscow Region, 141700, Russia*

[3]*Institute of Spectroscopy, Russian Academy of Sciences, Troitsk, Moscow, 108840, Russia*

[4]*Department of Physics, University of Bath, BA2 7AY Bath, UK*

*\*Corresponding author: zhangyiqi@mail.xjtu.edu.cn*

**Abstract:** We introduce a class of topological lasers based on the photonic Floquet topological insulator concept. The proposed system is realized as a truncated array of the lasing helical waveguides, where the pseudo-magnetic field arises due to twisting of the waveguides along the propagation direction that breaks the time-reversal symmetry and opens up a topological gap. When sufficient gain is provided in the edge channels of the array then the system lases into the topological edge states. Topological lasing is stable only in certain intervals of the Bloch momenta, that ensure a dynamic, but stable balance between the linear amplification and nonlinear absorption leading to the formation of the breathing edge states. We also illustrate topological robustness of the edge currents by simulating lattice defects and triangular arrangements of the waveguides.

An important property of topological insulators is the existence of the topologically protected states at its edges with energies inside a topological gap and connecting two bands with different topological invariants. In real space such edge states may demonstrate unidirectional propagation and topological robustness to the lattice and edge distortions[1,2]. Originated in solid-state physics, the concept of topological insulators is now interdisciplinary: It has been introduced also in mechanical[3], acoustic[4,5], atomic[6-9], photonic[10-24], optoelectronic[25-29] and many other systems, where diverse potential applications of topologically protected transport are envisioned. Recent progress in the sub-area of photonic topological insulators is described in, e.g., reviews[30,31].

Floquet topological insulators is a special case in a family of photonic realizations of topological systems, where a system is also periodic in an evolution variable, which can be either time or a longitudinal coordinate. Following the first proposal of such a system in semiconductor quantum wells[12], Floquet topological insulators have been realized with honeycomb arrays of helical waveguides[18]. In the latter case, the waveguide helicity gives rise to the pseudo-magnetic field breaking the time-reversal symmetry and leading to the appearance of the unidirectional edge states. Helical waveguide array is a photonic analogue of the Haldane system[32,33] in high-frequency driving limit, and it can be used to verify the anomalous quantum

Hall effect. A variety of new phenomena were theoretical predicted or experimentally observed with helical waveguide arrays, including anomalous topological insulators[19,20], topologically protected path entanglement[34], unpaired Dirac cones[35], topological edge states in quasi-crystals[21], solitons[36-38], topological Anderson insulator[22,39,40], topological phases in synthetic dimensions[24], guiding light by artificial gauge fields[41], and others. Notice that driven topological systems, such as helical arrays, may be characterized by special topological invariants[42].

Topological phases of matter are nowadays under active investigation not only in conservative, but also in dissipative settings, see for example[43-46]. Among the most exciting opportunities in this direction is the realization of lasing in topological edge states in active systems that promise remarkable stability of topological lasers, inherited from robustness and resistance to disorder of conservative topological systems. Theoretically topological lasers were proposed in photonic crystals[47]. Later they were realized in one-dimensional polaritonic and photonic structures employing Su-Schrieffer-Heeger model[48-52], which, however, did not allow to demonstrate topological currents due to the reduced dimensionality. Two-dimensional topological lasing was very recently observed in photonic crystals[53], lattices of coupled-ring resonators[54,55], and proposed theoretically in polaritonic arrays[56]. In these static systems, edge states appear either due to the external magnetic[53,56], or due to judicious engineering of coupling between elements leading to Haldane model[54,55]. At the same time, our proposal offers advantages of not using external magnetic fields, operating at optical frequencies, and relying on conventional nonlinear transparent materials.

The aim of this work is to show that such Floquet topological lasers exhibiting stable disorder- and defect-immune lasing in topologically protected edge states can be implemented using truncated honeycomb array of helical waveguides written or fabricated in the nonlinear optical material with gain saturation. Broken time-reversal symmetry guarantees the existence of unidirectional edge states that can lase, when spatially-inhomogeneous gain is provided for them. Nonlinear losses result in stabilization of the nonlinear edge states at certain Bloch momenta determining their group velocity.

We describe dynamics of light in Floquet topological lasers using the nonlinear Schrödinger equation for field amplitude that in dimensionless units takes the form:

$$i\frac{\partial \psi}{\partial z} = -\frac{1}{2}\nabla^2\psi - [\mathcal{R}_{\mathrm{re}}(x,y,z) - i\mathcal{R}_{\mathrm{im}}(x,y,z)]\psi - |\psi|^2\psi - i\gamma\psi - i\alpha|\psi|^2\psi, \qquad (1)$$

where $\psi$ is the scaled field amplitude; $x,y$ are the transverse coordinates normalized to the characteristic transverse scale $w_0$; $z$ is the propagation distance scaled to diffraction length $\kappa w_0^2$; $\kappa = 2\pi n_{\mathrm{r}}/\lambda$ is the wavenumber; $n_{\mathrm{r}}$ is the real part of the unperturbed refractive index of the material; $\gamma = \kappa^2 w_0^2 n_{\mathrm{i}}/n_{\mathrm{r}}$ is the scaled linear loss that is assumed uniform; $n_{\mathrm{i}} \ll n_{\mathrm{r}}$ is the imaginary part of the refractive index; $\alpha$ is the nonlinear loss parameter. Further we consider focusing cubic (Kerr) nonlinearity, typical for many solid materials, including optical fibres, but Floquet laser can be realized in the defocusing case too. We assume that the Floquet laser is composed from honeycomb array of helical waveguides that modulates the linear refractive index $\mathcal{R}_{\mathrm{re}}(x,y,z) = p_{\mathrm{re}}\sum_{n,m}\mathcal{Q}(x'-x_n, y'-y_m)$, where $p_{\text{re}} = \kappa^2 w_0^2 \delta n_{\text{r}} / n_{\text{r}}$ is the modulation depth, $x_n, y_m$ are the nodes of the honeycomb grid; $x' = x - r_0 \sin(\omega z)$ and $y' = y + r_0 - r_0 \cos(\omega z)$, where $r_0$ is the helix radius, $Z = 2\pi / \omega$ is the helix period, and $\mathcal{Q} = \exp[-(x^2 + y^2)^2 / d^4]$ is the function describing profile of individual waveguides of width $d$ [see Fig. 1(a) with schematic array representation and Fig. 4(c)]. The separation between the waveguides in the array is $a$. We assume that the array is truncated along the $x$-axis to form two zig-zag edges and that gain is provided only on its left edge $\mathcal{R}_{\text{im}}(x, y, z) = p_{\text{im}} \sum_{q,l} \mathcal{Q}(x' - x_q, y' - y_l)$, where $x_q, y_l$ are the coordinates of edge waveguides, see green waveguides in Fig. 1(a), while $p_{\text{im}} \ll p_{\text{re}}$ is the gain amplitude. The array is periodic in $y$ with period $Y = 3^{1/2} a$. Results do not change qualitatively for $z$-independent gain acting only inside the edge waveguides. It should be noted that by moving into the coordinate frame co-rotating with the waveguides $x \to x'$, $y \to y'$ Eq. (1) can be rewritten in the form:

$$i \frac{\partial \psi}{\partial z'} = -\frac{1}{2}[\nabla + i\mathbf{A}(z')]^2 \psi - [\mathcal{R}_{\text{re}}(x', y') - i\mathcal{R}_{\text{im}}(x', y')]\psi - \frac{1}{2} r_0^2 \omega^2 \psi - |\psi|^2 \psi - i\gamma\psi - i\alpha|\psi|^2 \psi, \quad (2)$$

where $\mathbf{A} = r_0 \omega[-\cos(\omega z'), \sin(\omega z')]$ is the gauge potential, $\mathcal{R}_{\text{re,im}}$ do not depend on $z' = z$. Further we select parameters of helical waveguide array in accordance with recent experiments[18], see caption to Fig. 1.

Dissipative helical arrays exhibiting gain in certain waveguides can be fabricated in different ways. Most tried approach relies on the direct laser writing with femtosecond pulses available in a broad range of transparent materials[57], including those containing amplifying dopants. Thus, various wave-guides were already realized in Er-doped active phosphate[58], silicate [59], tellurite [60], and Baccarat[61] glasses, and also in lithium niobate[62] allowing realization of inhomogeneous parametric gain used for observation of parity-time symmetry[63]. Another viable alternative is the infiltration of hollow photonic crystal fibres with helical channels with active index-matching liquids[64,65].

First, we consider topological properties of conservative linear helical waveguide array by setting $\gamma, \alpha, p_{\text{im}} = 0$ and neglecting nonlinearity in Eq. (1). The eigenstates $\psi(x,y,z) = u(x,y,z)\exp(i\beta z + iky)$ of Eq. (1) are Bloch waves, where $u$ is localized in $x$: $u_{x \to \pm\infty} \to 0$ and periodic both in $y$ and $z$ directions: $u_{z+Z} = u_z$ and $u_{y+Y} = u_y$, $k$ is the Bloch momentum along the $y$-axis, $\beta$ is the quasienergy. The latter is periodic function of $k$ with period $K = 2\pi / Y$ and is defined modulo $\omega = 2\pi / Z$ due to longitudinal periodicity of the array. Typical quasienergy spectrum for helical array is presented in Fig. 1(b). Since we consider real-world continuous system, quasienergy spectrum was calculated using following approach. First, Bloch modes $\psi_i^{\text{st}} = u_i^{\text{st}}(x,y)\exp(i\beta z + iky)$ from two top bands of the static truncated array with straight channels were obtained using a plane-wave expansion method. The number of such modes is $2n$, where $n$ is the number of waveguides in one $y$-period of the array (unit cell). Each such mode $\psi_i^{\text{st}}$, normalized as $(\psi_i^{\text{st}}, \psi_j^{\text{st}}) = \delta_{ij}$, where Hermitian product involves integral over one unit cell of the array, was propagated in helical array for one period Z. Rotation couples modes from the first two bands (coupling to the lower bands can be neglected, since they remain well-separated). The output distributions $\psi_j^{\text{out}}$ corresponding to input $\psi_j^{\text{st}}$ were then projected on the initial basis of modes $\psi_i^{\text{st}}$ that yields $2n \times 2n$ projection matrix $\mathcal{H}_{ij} = (\psi_i^{\text{st}}, \psi_j^{\text{out}})$, whose eigenvalues are Floquet exponents $\exp(i\lambda_j)$. Quasienergies are found as $\beta_j = \lambda_j / Z$; their imaginary part is negligible as long as radiative losses are small that is the case for parameters used below.

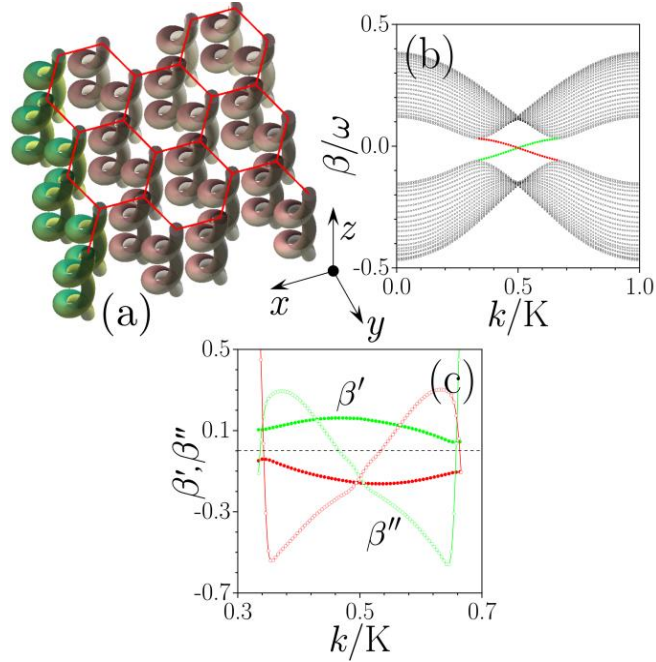

Fig. 1. (a) Schematic illustration of truncated helical waveguide array. Amplifying edge waveguides are shown green, waveguides with losses in the bulk are shown brown. (b) Quasienergies $\beta$ of linear modes supported by conservative truncated helical array versus Bloch momentum $k$. (c) Velocity $\beta'$ (solid circles) and dispersion $\beta''$ (open circles) of the edge states versus Bloch momentum $k$. Red (green) circles correspond to the edge states from the left (right) edges of the array, black circles correspond to bulk modes. Here and below helix radius $r_0 = 0.5$, period $Z = 6$, waveguide width $d = 0.4$ and separation $a = 1.6$, array depth $p_{\text{re}} = 8.9$.

Waveguide rotation opens topological gap with the edge states existing for $\mathrm{K}/3 < k < 2\mathrm{K}/3$ and the zigzag-zigzag interface [Fig. 1(b)]. The width of the gap increases with increase of helix radius $r_0$ or decrease of rotation period $Z$, but so do also radiative losses, so for each $Z$ there is certain optimal $r_0$. There are two topological edge states in the gap – red curve corresponds to the states on the left edge that moves in positive $y$-direction, green curve corresponds to the right edge states moving in the negative $y$-direction. Black circles correspond to bulk modes. First $\beta' = \partial\beta/\partial k$ and second $\beta'' = \partial^2\beta/\partial k^2$ derivatives of the quasienergy that quantify group velocity and dispersion of the edge states are shown in Fig. 1(c). Inversion of the waveguide rotation direction also inverts the direction of edge currents. The sign and magnitude of $\beta''$ determines domains of Bloch momentum, where modulational instability of the edge state can develop in the presence of nonlinearity in a conservative case (thus, for focusing nonlinearity this is possible when $\beta'' < 0$ [38]), but in a dissipative system such instabilities may be suppressed by linear and nonlinear losses, as shown below. An array with the bearded edges can be analysed similarly and the edge states were found for $k < \mathrm{K}/3$ and $k > 2\mathrm{K}/3$.

Our system retains its topological properties in the presence of the spatially uniform linear losses $\gamma$ and gain $p_{\mathrm{im}}$ concentrated in the edge channels. The edge states are well localized, and hence they have largest overlap with gain area relative to the other modes and therefore they experience preferential amplification. We have found that there exists a sharp threshold in $p_{\mathrm{im}}$ above which lasing in edge states occurs. Most efficient amplification occurs for the edge states with Bloch momentum, which most localised around the edge. The interval of $k$ values, where the edge states get amplified increases with $p_{\mathrm{im}}$ until lasing becomes possible in the entire topological gap. Because gain was provided on the left edge only, the right edge states were attenuated.

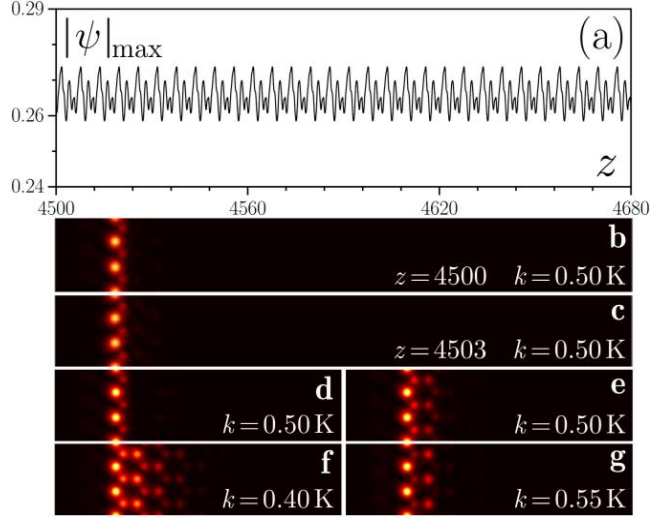

Fig. 2. (a) Typical evolution of peak amplitude in stable nonlinear edge state in Floquet laser at $p_{\text{im}} = 0.13$, $k = 0.5\text{K}$ and (b),(c) representative field modulus distributions at $z = n\text{Z}$ and $z = (n+1/2)\text{Z}$. Examples of the nonlinear edge states at $k = 0.5\text{K}$, $p_{\text{im}} = 0.15$ (d) and $p_{\text{im}} = 0.21$ (e); $k = 0.4\text{K}$, $p_{\text{im}} = 0.15$ (f); and $k = 0.55\text{K}$, $p_{\text{im}} = 0.15$ (g). The edge states in (b)-(d) are stable, while edge states in (e)-(g) are unstable. In all cases $\alpha = 0.5$.

To achieve stable lasing we now add focusing nonlinearity and nonlinear absorption into system. This leads to appearance of attractors – nonlinear edge states performing periodic stable breathing in the course of propagation. A typical breathing dynamics of the edge state is illustrated in Figs. 2(a)-2(c). Being stable attractor, this state was excited using a linear conservative state with $k = \text{K}/2$ as an initial condition. After some transient stage the state has evolved into completely stable nonlinear dissipative mode existing due to balance between nonlinearity and diffraction, gain and losses, that exactly replicates its transverse profile after each period $\text{Z}$. The amplitude of this nonlinear edge state shows complex, but regular periodic oscillations without damping or growth and the period of these oscillations coincides with helix period $Z=6$. Figure 2(a) shows thirty of these periods to stress that the state is practically stable. Amplitude oscillations notably increase with the increasing gain $p_{\text{im}}$. Comparison of wave profiles at $z=nZ$ [Fig. 2(b)] and $z=(n+1/2)Z$ [Fig. 2(c)] shows slightly larger penetration of the latter state into the depth of array. Increasing gain amplitude leads to gradual expansion of the dissipative state into the depth of array [Figs. 2(d) and 2(e)] and may finally cause its destabilization [Fig. 2(d) shows a state from the boundary of the stability domain, while the state in Fig. 2(e) is unstable]. The extend of the edge states away from the edge and into the crystal strongly depends on Bloch momentum $k$ and increases for the quasienergies approaching the gap boundaries [Figs. 2(f) and 2(g)].

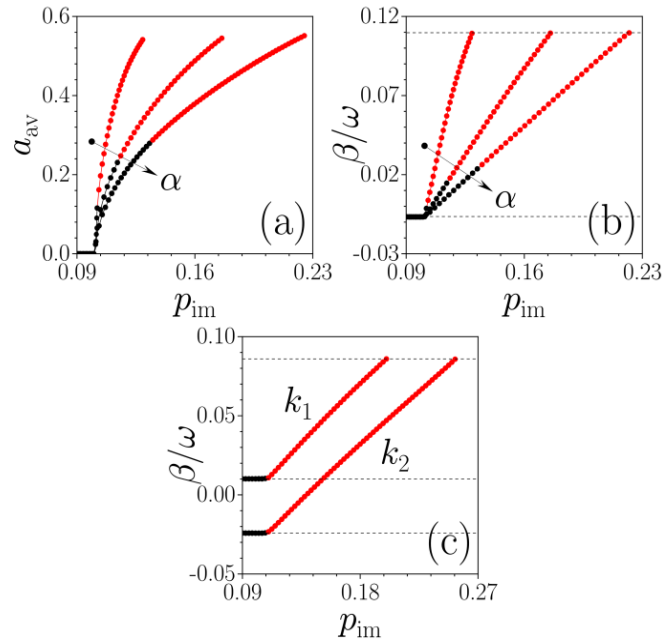

Fig. 3. Average amplitude (a) and quasienergy (b) of the nonlinear edge state versus gain amplitude $p_{\text{im}}$ at $k=0.5K$ for the different values of nonlinear absorption, $\alpha=0.1$, $0.2$, and $0.3$ (the direction of increase of $\alpha$ is shown by arrows). Lower and upper dashed lines in (b) indicate quasienergy of lin-

ear edge state and the border of topological gap for $k = 0.5\text{K}$. (c) Quasienergy of the nonlinear edge state versus gain amplitude $p_{\text{im}}$ for $k_1 = 0.45\text{K}$ and $k_2 = 0.55\text{K}$, at $\alpha = 0.5$. Upper dashed line indicates the border of the gap, identical for $k_1$ and $k_2$, while two lower dashed lines indicate energies of linear edge states, which are different for $k_1$ and $k_2$. Stable branches are shown black, unstable branches are shown red.

To prove that nonlinear dissipative edge states reported here are topological, we traced families of such states by gradually increasing gain amplitude $p_{\text{im}}$. Since amplitude of the lasing state exhibits complex behaviour over one helix period [see Fig. 2(a)], we introduce a new quantity – averaged amplitude $a_{\text{av}} = \text{Z}^{-1} \int_{n\text{Z}}^{(n+1)\text{Z}} |\psi|_{\max} dz$. This amplitude is depicted in Fig. 3(a) as a function of $p_{\text{im}}$ for $k = \text{K}/2$ and different values of the nonlinear absorption coefficient $\alpha$. Moreover, we introduced quasienergy $\beta$ of the nonlinear dissipative edge states by analogy with quasienergy of linear conservative states. It can be determined from phase $\phi$ accumulated by the edge state over one helix period $\beta = \phi/\text{Z}$, where phase $\phi$ is calculated numerically from the product $(\psi_{z=n\text{Z}}, \psi_{z=(n+1)\text{Z}}) = U \exp(i\phi)$, where $U$ is the norm of the state per one unit cell ($y$-period). The dependencies $\beta(p_{\text{im}})$, calculated for the same values of nonlinear absorption $\alpha$ as in $a_{\text{av}}(p_{\text{im}})$ curves, are shown in Fig. 3(b). The presence of lasing threshold in $p_{\text{im}}$ is obvious in Fig. 3(a) – it corresponds to the point where averaged amplitude of the edge state becomes nonzero. Lasing threshold is minimal for Bloch momentum $k = \text{K}/2$ (in this case $p_{\text{im}} \approx 0.1$) and it slightly increases for other momentum values reaching maximal values at $k \to \text{K}/3$ or $k \to 2\text{K}/3$, the property connected with decreasing overlap of the edge states for latter momentum values with gain landscape leading to less ef-ficient amplification. The quasienergy $\beta$ of the nonlinear edge state at its generation threshold coincides with that of the conservative linear state, as indicated by the bottom dashed line in Fig. 3(b), and it increases almost linearly with increasing gain until it reaches the upper edge of the topological gap, as indicated by the top dashed line in Fig. 3(b). Thus nonlinear edge states bifurcate from linear ones, once gain $p_{\rm im}$ exceeds corresponding $k$-dependent threshold. This is illustrated in Fig. 3(c), where dependencies $\beta(p_{\rm im})$ for different momentum values $k_1 = 0.45\,{\rm K}$ and $k_2 = 0.55\,{\rm K}$ clearly start at different levels coinciding with quasienergies of corresponding linear edge states from red branch of Fig. 1(b). If quasienergy of the nonlinear edge state moves out of the topological gap for a given $k$, this state acquires nonzero background inside the array due to coupling to bulk modes, thus we truncate the $a_{\rm av}(p_{\rm im})$ and $\beta(p_{\rm im})$ dependences in Figs. 3 accordingly. Notice that in Fig. 3(c) the upper edge of the gap is the same for two presented $k$ values.

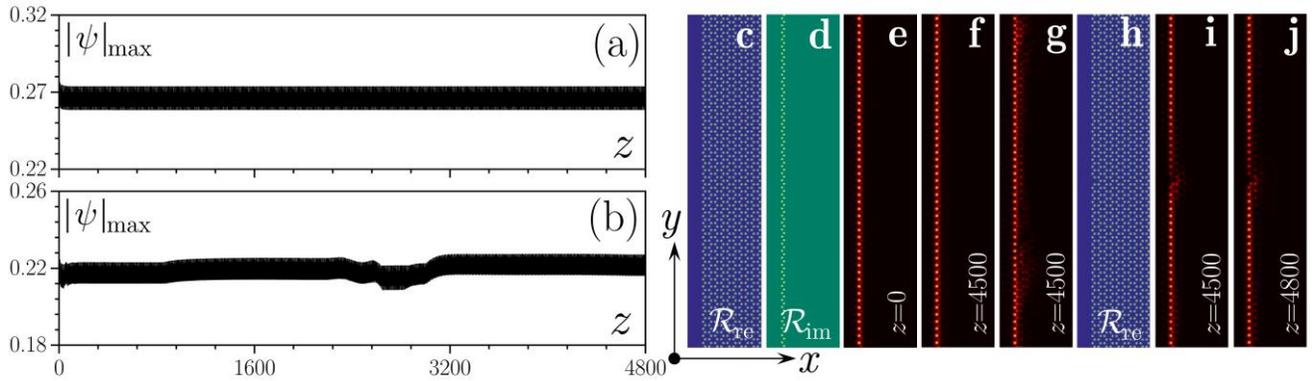

Fig. 4. Peak amplitude versus distance illustrating (a) stable propagation of the edge state at $p_{\rm im} = 0.13$ in regular Floquet laser (c),(d) [corresponding $|\psi|$ distributions are shown in (e),(f)]; and (b) stable propagation in the presence of edge defect in the form of missing channel (h) at $p_{\rm im} = 0.12$ [corresponding $|\psi|$ distributions at different times are shown in (i),(j)]. In-stability development is shown in (g) for edge state with $p_{\text{im}} = 0.2$. In all cases $k = 0.5\text{K}$, $\alpha = 0.5$ and input states were perturbed by $5\%$ amplitude noise

We also tested stability of all obtained dissipative edge states by perturbing them with $5\%$ amplitude noise and modelling their long-distance propagation on huge transverse windows (100 $y$-periods) to accommodate for all possible perturbations that could lead to instability of these states. The outcome is that for $k = \text{K}/2$ considerable portions of branches of nonlinear states close to lasing threshold are stable. In Fig. 3 stable families are marked with black dots, while the unstable ones are marked with red dots. Two observations can be made: Increasing gain eventually leads to destabilisation of the nonlinear states, but higher nonlinear absorption extends stability intervals, Fig. 3(a). The interval of stability in gain amplitudes $p_{\text{im}}$ quickly decreases away from momentum $k = \text{K}/2$, so that all states with momenta $k = 0.45\,\text{K}$ and $k = 0.55\,\text{K}$ were found formally unstable. However, corresponding instabilities are very weak, so in practical experimental conditions with finite samples and close to lasing threshold such states will appear as stable ones too.

Stable propagation of the perturbed dissipative edge state in Floquet laser is illustrated in Figs. 4(a),(e),(f) for $k = \text{K}/2$. In Fig. 4(a), the peak amplitude $|\psi|_{\max}$ of the launched state is shown during propagation that clearly performs regular periodic oscillations reflecting helical structure of the waveguide array. Note that the curve in Fig. 2(a) is a portion (in the range $4500 \leq z \leq 4680$) of dependence in Fig. 4(a). Comparison of initial and output field modulus distributions in Figs. 4(e) and 4(f) showing only small fraction of actual integration window in $y$, reveals complete stability of the wave. In contrast, development of instability is shown in Fig. 4(g) for large gain amplitude $p_{\text{im}} = 0.2$. Even in this case, despite the appearance of weak irregular modulations travelling along array interface and weak radiation into the bulk, the state remains confined near the interface at any propagation distance.

The striking advantage of dissipative topological edge states in Floquet laser is that they inherit topological protection of conservative edge states. To illustrate this we remove one waveguide from the left edge of helical waveguide array. The real part of corresponding array is depicted in Fig. 4(h); there is similar defect in gain profile too (not shown). Dissipative edge state launched into such a helical waveguide array at $p_{\text{im}}=0.12$ experience some reshaping and amplitude oscillations due to presence of defect [see Fig. 4(b)], but finally reaches new stationary state shown in Figs. 4(i) and 4(j) for different distances. The representative feature of these distributions is that state is perturbed only locally around the defect and no radiation into bulk is visible.

Finally, we note that practical Floquet lasers should be spatially compact, hence we considered also triangular geometry of the array, in which gain is again provided only in edge channels, see Figs. 5(c) and 5(d). Such a geometry may be beneficial for formation of stable edge currents, because it allows to effectively eliminate instabilities to low-frequency perturbations. To simultaneously illustrate edge currents and formation of stable attractor in this system, we start with localized excitation on the left edge of the triangle with broad Gaussian envelope [Fig. 5(e)] and let it evolve at $p_{\text{im}}=0.13$ and $\alpha=0.5$. Figures 5(f)-5(h) reveal clockwise circulation of the state accompanied by its gradual expansion only along the edge of array. As mentioned above, inversion of the rotation direction of waveguides inverts also the direction of edge current in this system. Already at distances $z\sim 500$ the entire edge of the array becomes excited. After some transient stage the wave reaches steady-state profile depicted in Fig. 5(h), while peak amplitude of the wave stops changing [Fig. 5(a)]. Notice excellent lo-calization near the edge of the structure. To verify the topological protection of states in this finite system we introduced two defects to the triangular insulator by removing two channels from top and bottom edges; the corresponding real part of the array is shown in Fig. 5(i). As for the gain landscape, we removed only one channel on the bottom edge, but kept corresponding channel on the top one (not shown here). Using the same initial excitation as in Fig. 5(e) we arrived to the final steady-state profile shown in Fig. 5(j) that exhibits local deformations only around defect channels. In the presence of defects steady-state regime is reached at somewhat larger propagation distances [see Fig. 5(b) with corresponding dependence of peak amplitude on $z$]. Interestingly, deformed patterns on the top and bottom edges look practically the same, that indicates that the particular type of the defect (purely conservative or dissipative) is not important due to topological protection in Floquet laser

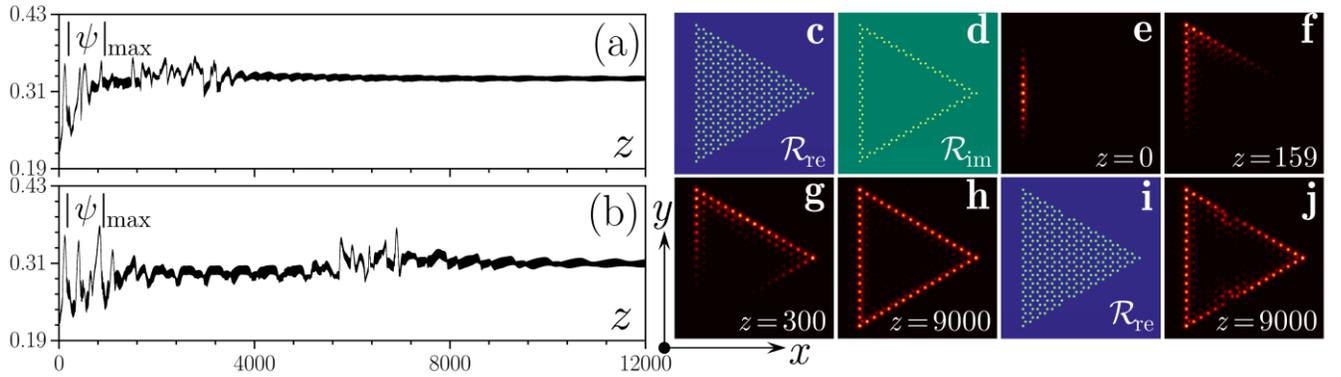

Fig. 5. Peak amplitude versus distance illustrating stable circulation in triangular Floquet laser without edge defects (a) and with edge defects (b) at $p_{\mathrm{im}} = 0.13$, $\alpha = 0.5$. Refractive index (c) and gain (d) distributions in Floquet laser without defects and $|\psi|$ snapshots (e)-(h) illustrating circulation in this structure. Refractive index (i) and $|\psi|$ distribution at large distance (j) in the Floquet laser with two edge defects

Summarizing, we have investigated the topological lasing in photonic Floquet topological insulators. We demonstrated that the edge states in this system are topologically protected against large structural perturbations and can be either dynamically stable or unstable depending on the system parameters and, in particular, on the gain amplitude. We demonstrated lasing not only for an idealized infinite edge but also for a more practical triangular geometry. This work provides a practically feasible scheme to obtain topological lasing without the external magnetic fields.

This work was supported by Natural Science Foundation of Guangdong province (2018A0303130057), Fundamental Research Funds for the Central Universities (xzy012019038, xzy022019076), and RFBR and DFG according to the research project № 18-502-12080.

**References**

1. M. Z. Hasan and C. L. Kane, "Colloquium: Topological insulators," Rev. Mod. Phys. **82**, 3045-3067 (2010).

2. X.-L. Qi and S.-C. Zhang, "Topological insulators and superconductors," Rev. Mod. Phys. **83**, 1057 (2011).

3. S. D. Huber, "Topological mechanics," Nat. Phys. **12**, 621-623 (2016).

4. Z. Yang, F. Gao, X. Shi, X. Lin, Z. Gao, Y. Chong, and B. Zhang, "Topological acoustics," Phys. Rev. Lett. **114**, 114301 (2015).

5. C. He, X. Ni, H. Ge, X.-C. Sun, Y.-B. Chen, M.-H. Lu, X.-P. Liu, and Y.-F. Chen, "Acoustic topological insulator and robust one-way sound transport," Nat. Phys. **12**, 1124-1129 (2016).

6. G. Jotzu, M. Messer, R. Desbuquois, M. Lebrat, T. Uehlinger, D. Greif, and T. Esslinger, "Experimental realization of the topological Haldane model with ultracold fermions," Nature **515**, 237-240 (2014).

7. M. Aidelsburger, M. Lohse, C. Schweizer, M. Atala, J. T. Barreiro, S. Nascimbène, N. R. Cooper, I. Bloch, and N. Goldman, "Measuring the Chern number of Hofstadter bands with ultracold bosonic atoms," Nat. Phys. **11**, 162-166 (2015).

8. M. C. Beeler, R. A. Williams, K. Jimenez-Garcia, L. J. LeBlanc, A. R. Perry, and I. B. Spielman, "The spin Hall effect in a quantum gas," Nature **498**, 201-204 (2013).

9. C. J. Kennedy, G. A. Siviloglou, H. Miyake, W. C. Burton, and W. Ketterle, "Spin-orbit coupling and quantum spin Hall effect for neutral atoms without spin flips," Phys. Rev. Lett. **111**, 225301 (2013).

10. F. D. M. Haldane and S. Raghu, "Possible Realization of Directional Optical Waveguides in Photonic Crystals with Broken Time-Reversal Symmetry," Phys. Rev. Lett. **100**, 013904 (2008).

11. Z. Wang, Y. Chong, J. D. Joannopoulos, and M. Soljacic, "Observation of unidirectional backscattering-immune topological electromagnetic states," Nature **461**, 772-775 (2009).

12. N. H. Lindner, G. Refael, and V. Galitski, "Floquet topological insulator in semiconductor quantum wells," Nat. Phys. **7**, 490-495 (2011).

13. M. Hafezi, E. A. Demler, M. D. Lukin, and J. M. Taylor, "Robust optical delay lines with topological protection," Nat. Phys. **7**, 907-912 (2011).

14. R. O. Umucalilar and I. Carusotto, "Fractional Quantum Hall States of Photons in an Array of Dissipative Coupled Cavities," Phys. Rev. Lett. **108**, 206809 (2012).

15. M. Hafezi, S. Mittal, J. Fan, A. Migdall, and J. M. Taylor, "Imaging topological edge states in silicon photonics," Nat. Photon. **7**, 1001-1005 (2013).

16. A. B. Khanikaev, S. H. Mousavi, W.-K. Tse, M. Kargarian, A. H. MacDonald, and G. Shvets, "Photonic topological insulators," Nat. Mater. **12**, 233-239 (2013).

17. W.-J. Chen, S.-J. Jiang, X.-D. Chen, B. Zhu, L. Zhou, J.-W. Dong, and C. T. Chan, "Experimental realization of photonic topological insulator in a uniaxial metacrystal waveguide," Nat. Commun. **5**, 5782 (2014).

18. M. C. Rechtsman, J. M. Zeuner, Y. Plotnik, Y. Lumer, D. Podolsky, F. Dreisow, S. Nolte, M. Segev, and A. Szameit, "Photonic Floquet topological insulators," Nature **496**, 196-200 (2013).

19. L. J. Maczewsky, J. M. Zeuner, S. Nolte, and A. Szameit, "Observation of photonic anomalous Floquet topological insulators," Nat. Commun. **8**, 13756 (2017).

20. S. Mukherjee, A. Spracklen, M. Valiente, E. Andersson, P. Öhberg, N. Goldman, and R. R. Thomson, "Experimental observation of anomalous topological edge modes in a slowly driven photonic lattice," Nat. Commun. **8**, 13918 (2017).

21. M. A. Bandres, M. C. Rechtsman, and M. Segev, "Topological photonic quasicrystals: Fractal topological spectrum and protected transport," Phys. Rev. X **6**, 011016 (2016).

22. S. Stützer, Y. Plotnik, Y. Lumer, P. Titum, N. H. Lindner, M. Segev, M. C. Rechtsman, and A. Szameit, "Photonic topological Anderson insulators," Nature **560**, 461-465 (2018).

23. Y. Yang, Z. Gao, H. Xue, L. Zhang, M. He, Z. Yang, R. Singh, Y. Chong, B. Zhang, and H. Chen, "Realization of a three-dimensional photonic topological insulator," Nature **565**, 622-626 (2019).

24. E. Lustig, S. Weimann, Y. Plotnik, Y. Lumer, M. A. Bandres, A. Szameit, and M. Segev, "Photonic topological insulator in synthetic dimensions," Nature **567**, 356-360 (2019).

25. A. V. Nalitov, D. D. Solnyshkov, and G. Malpuech, "Polariton Z Topological Insulator," Phys. Rev. Lett. **114**, 116401 (2015).

26. T. Karzig, C.-E. Bardyn, N. H. Lindner, and G. Refael, "Topological Polaritons," Phys. Rev. X **5**, 031001 (2015).

27. Y. V. Kartashov and D. V. Skryabin, "Modulational instability and solitary waves in polariton topological insulators," Optica **3**, 1228-1236 (2016).

28. C. Li, F. Ye, X. Chen, Y. V. Kartashov, A. Ferrando, L. Torner, and D. V. Skryabin, "Lieb polariton topological insulators," Phys. Rev. B **97**, 081103(R) (2018).

29. S. Klembt, T. H. Harder, O. A. Egorov, K. Winkler, R. Ge, M. A. Bandres, M. Emmerling, L. Worschech, T. C. H. Liew, M. Segev, C. Schneider, and S. Höfling, "Exciton-polariton topological insulator," Nature **562**, 552-556 (2018).

30. L. Lu, J. D. Joannopoulos, and M. Soljačić, "Topological photonics," Nat. Photon. **8**, 821-829 (2014).

31. T. Ozawa, H. M. Price, A. Amo, N. Goldman, M. Hafezi, L. Lu, M. Rechtsman, D. Schuster, J. Simon, O. Zilberberg, and I. Carusotto, "Topological photonics," Rev. Mod. Phys. **91**, 015006 (2019).

32. F. D. M. Haldane, "Model for a Quantum Hall Effect without Landau Levels: Condensed-Matter Realization of the 'Parity Anomaly'," Phys. Rev. Lett. **61**, 2015-2018 (1988).

33. D.-W. Wang, H. Cai, L. Yuan, S.-Y. Zhu, and R.-B. Liu, "Topological phase transitions in superradiance lattices," Optica **2**, 712-715 (2015).

34. M. C. Rechtsman, Y. Lumer, Y. Plotnik, A. Perez-Leija, A. Szameit, and M. Segev, "Topological protection of photonic path entanglement," Optica **3**, 925-930 (2016).

35. D. Leykam, M. C. Rechtsman, and Y. D. Chong, "Anomalous Topological Phases and Unpaired Dirac Cones in Photonic Floquet Topological Insulators," Phys. Rev. Lett. **117**, 013902 (2016).

36. M. J. Ablowitz, C. W. Curtis, and Y.-P. Ma, "Linear and nonlinear traveling edge waves in optical honeycomb lattices," Phys. Rev. A **90**, 023813 (2014).

37. Y. Lumer, Y. Plotnik, M. C. Rechtsman, and M. Segev, "Self-Localized States in Photonic Topological Insulators," Phys. Rev. Lett. **111**, 243905 (2013).

38. D. Leykam and Y. D. Chong, "Edge Solitons in Nonlinear-Photonic Topological Insulators," Phys. Rev. Lett. **117**, 143901 (2016).

39. P. Titum, N. H. Lindner, M. C. Rechtsman, and G. Refael, "Disorder-Induced Floquet Topological Insulators," Phys. Rev. Lett. **114**, 056801 (2015).

40. P. Titum, E. Berg, M. S. Rudner, G. Refael, and N. H. Lindner, "Anomalous Floquet-Anderson Insulator as a Nonadiabatic Quantized Charge Pump," Phys. Rev. X **6**, 021013 (2016).

41. Y. Lumer, M. A. Bandres, M. Heinrich, L. J. Maczewsky, H. Herzig-Sheinfux, A. Szameit, and M. Segev, "Light guiding by artificial gauge fields," Nat. Photon. **13**, 339-345 (2019).

42. M. S. Rudner, N. H. Lindner, E. Berg, and M. Levin, "Anomalous Edge States and the Bulk-Edge Correspondence for Periodically Driven Two-Dimensional Systems," Phys. Rev. X **3**, 031005 (2013).

43. H. Schomerus, "Topologically protected midgap states in complex photonic lattices," Opt. Lett. **38**, 1912-1914 (2013).

44. J. M. Zeuner, M. C. Rechtsman, Y. Plotnik, Y. Lumer, S. Nolte, M. S. Rudner, M. Segev, and A. Szameit, "Observation of a Topological Transition in the Bulk of a Non-Hermitian System," Phys. Rev. Lett. **115**, 040402 (2015).

45. S. Weimann, M. Kremer, Y. Plotnik, Y. Lumer, S. Nolte, K. G. Makris, M. Segev, M. C. Rechtsman, and A. Szameit, "Topologically protected bound states in photonic parity–time-symmetric crystals," Nat. Mater. **16**, 433-438 (2017).

46. D. Leykam, K. Y. Bliokh, C. Huang, Y. D. Chong, and F. Nori, "Edge Modes, Degeneracies, and Topological Numbers in Non-Hermitian Systems," Phys. Rev. Lett. **118**, 040401 (2017).

47. L. Pilozzi and C. Conti, Topological lasing in resonant photonic structures, Phys. Rev. B **93**, 195317 (2016).

48. P. St-Jean, V. Goblot, E. Galopin, A. Lemaítre, T. Ozawa, L. Le Gratiet, I. Sagnes, J. Bloch, and A. Amo, "Lasing in topological edge states of a 1D lattice," Nat. Photon. **11**, 651-656 (2017).

49. M. Parto, S. Wittek, H. Hodaei, G. Harari, M. A. Bandres, J. Ren, M. C. Rechtsman, M. Segev, D. N. Christodoulides, and M. Khajavikhan, "Edge-Mode Lasing in 1D Topological Active Arrays," Phys. Rev. Lett. **120**, 113901 (2018).

50. H. Zhao, P. Miao, M. H. Teimourpour, S. Malzard, R. El-Ganainy, H. Schomerus, and L. Feng, Topological hybrid silicon microlasers, Nat. Commun. **9**, 981 (2018).

51. S. Longhi, Y. Kominis, and V. Kovanis, "Presence of temporal dynamical instabilities in topological insulator lasers," EPL **122**, 14004 (2018).

52. S. Malzard and H. Schomerus, "Nonlinear mode competition and symmetry-protected power oscillations in topological lasers," New J. Phys. **20**, 063044 (2018).

53. B. Bahari, A. Ndao, F. Vallini, A. El Amili, Y. Fainman, and B. Kanté, "Nonreciprocal lasing in topological cavities of arbitrary geometries," Science **358**, 636 (2017).

54. G. Harari, M. A. Bandres, Y. Lumer, M. C. Rechtsman, Y. D. Chong, M. Khajavikhan, D. N. Christodoulides, and M. Segev, "Topological insulator laser: Theory," Science **359**, eaar4003 (2018).

55. M. A. Bandres, S. Wittek, G. Harari, M. Parto, J. Ren, M. Segev, D. N. Christodoulides, and M. Khajavikhan, "Topological insulator laser: Experiment," Science **359**, eaar4005 (2018).

56. Y. V. Kartashov and D. V. Skryabin, "Two-Dimensional Topological Polariton Laser," Phys. Rev. Lett. **122**, 083902 (2019).

57. R. R. Gattass and E. Mazur, "Femtosecond laser micromachining in transparent materials," Nat. Photon. **2**, 219-225 (2008).

58. N. Chiodo, G. Della Valle, R. Osellame, S. Longhi, G. Cerullo, R. Ramponi, and P. Laporta, "Imaging of Bloch oscillations in erbium-doped curved waveguide arrays," Opt. Lett. **31**, 1651-1653 (2006).

59. R. R. Thomson, S. Campell, I. J. Blewett, A. K. Kar, D. T. Reid, S. Shen and A. Jha, "Active waveguide fabrication in erbium-doped oxyfluoride silicate glass using femtosecond pulses," Appl. Phys. Lett. **87**, 121102 (2005).

60. T. T. Fernandez, G. Della Valle, R. Osellame, G. Jose, N. Chiodo, A. Jha, and P. Laporta, "Active waveguides written by femtosecond laser irradiation in an erbium-doped phospho-tellurite glass," Opt. Express **16** 15198-15205 (2008).

61. K. C. Vishnubhatla, S. V. Rao, R. S. S. Kumar, R. Osellame, S. N. B. Bhaktha, S. Turrell, A. Chiappini, A. Chiasera, M. Ferrari, M. Mattarelli, M. Montagna, R. Ramponi, G. C. Righini, and D. N. Rao, "Femtosecond laser direct writing of gratings and waveguides in high quantum efficiency erbium-doped Baccarat glass," Phys. D: Appl. Phys. **42**, 205106 (2009).

62. J. Burghoff, C. Grebing, S. Nolte, and A. Tünnermann, "Efficient frequency doubling in femtosecond laser-written waveguides in lithium niobate," Appl. Phys. Lett. **89**, 081108 (2006).

63. C. E. Rüter, K. G. Makris, R. El-Ganainy, D. N. Christodoulides, M. Segev, and D. Kip, "Observation of parity–time symmetry in optics," Nat. Phys. **6**, 192-195 (2010).

64. M. Vieweg, T. Gissibl, S. Pricking, B. T. Kuhlmey, D. C. Wu, B. J. Eggleton, and H. Giessen, "Ultrafast nonlinear optofluidics in selectively liquid-filled photonic crystal fibers," Opt. Express **18**, 25232-25240 (2010).

65. M. Vieweg, S. Pricking, T. Gissibl, Y. V. Kartashov, L. Torner, and H. Giessen, "Tunable ultrafast nonlinear optofluidic coupler," Opt. Lett. **37**, 1058-1060 (2012).